\documentclass[10pt,conference]{IEEEtran}\IEEEoverridecommandlockouts
\usepackage{epsfig,graphicx,subfigure,psfrag,amsmath,cases}
\usepackage{latexsym,amssymb,amsmath,epsfig,subfigure,algorithm,mathtools}
\usepackage{algorithmic}
\usepackage{color}
\usepackage{url}
\usepackage{scrtime}
 \usepackage[left=0.6in,top=0.4in,bottom=0.47in,right=0.6in]{geometry}
\author{\authorblockN{Yan Sun\authorrefmark{1}, Derrick Wing Kwan Ng\authorrefmark{2}, Zhiguo Ding\authorrefmark{3}, and Robert Schober\authorrefmark{1}\thanks{Robert Schober is also with the University of British Columbia, Canada.}}

Institute for Digital Communications, Friedrich-Alexander-University
Erlangen-N\"urnberg (FAU), Germany\authorrefmark{1}\\
School of Electrical Engineering and Telecommunications, The University
of New South Wales, Australia\authorrefmark{2}\\
School of Computing and Communications, Lancaster University, United Kingdom\authorrefmark{3}\vspace*{-5mm}
}

\title{Optimal Joint Power and Subcarrier Allocation for MC-NOMA Systems}

\newtheorem{Def}{Definition}

\newtheorem{T-Prob}{Transformed Problem}

\DeclareMathOperator{\maxo}{maximize}
\DeclareMathOperator{\mino}{minimize}

\newcommand{\abs}[1]{\lvert#1\rvert}
\newcommand{\norm}[1]{\lVert#1\rVert}
\begin{document}
\maketitle

\begin{abstract}
In this paper, we investigate the resource allocation algorithm design for multicarrier non-orthogonal multiple access (MC-NOMA) systems. The proposed algorithm is obtained from the solution of a non-convex optimization problem for the maximization of the weighted system throughput.
We employ monotonic optimization to develop the optimal joint power and subcarrier allocation policy. The optimal resource allocation policy serves as a performance benchmark due to its high complexity.
Furthermore, to strike a balance between computational complexity and optimality, a suboptimal scheme with low computational complexity is proposed. Our simulation results reveal that the suboptimal algorithm achieves a close-to-optimal
performance and MC-NOMA employing the proposed resource allocation algorithm provides a substantial system throughput improvement compared to conventional multicarrier orthogonal multiple access (MC-OMA).
\end{abstract}
\renewcommand{\baselinestretch}{0.979}
\normalsize
\vspace*{-1mm}
\section{Introduction}
Multicarrier techniques have been widely adopted in broadband wireless communications over the last decade, due to their flexibility in resource allocation and their ability to exploit multiuser diversity \cite{Dynamic_FDrelay,Cui2009Distrib}. In conventional multicarrier systems, a given radio frequency band is divided into multiple subcarriers and each subcarrier is allocated to at most one user in order to avoid multiuser interference. Thus, spectral efficiency can be improved by performing user scheduling and power allocation.
In \cite{Dynamic_FDrelay}, the authors proposed an optimal joint precoding and scheduling algorithm for the maximization of the weighted system throughput in multiple-input multiple-output (MIMO) orthogonal frequency division multiple access (OFDM) full-duplex relaying systems.
The authors of \cite{Cui2009Distrib} proposed a distributed subcarrier, power, and rate allocation algorithm for the maximization of the weighted throughput in relay-assisted OFDM systems.
However, with the schemes in \cite{Dynamic_FDrelay,Cui2009Distrib}, the spectral resource is still underutilized as subcarriers may be assigned exclusively to a user with poor channel quality to ensure fairness.

Non-orthogonal multiple access (NOMA) has recently received significant attention since it enables the multiplexing of multiple users on the same frequency resource, which improves the system spectral efficiency \cite{ImpactPairNOMA}\nocite{ding2015general,minmaxNOMA,saito2013non,OFDM_NOMA_DC}--\cite{JointNOMA}. Since multiplexing multiple users on the same frequency channel leads to co-channel interference (CCI), successive interference cancellation (SIC) is performed at the receivers to remove the undesired interference.
The authors of \cite{ImpactPairNOMA} investigated the impact of user pairing on the sum rate of NOMA systems, and it was shown that the system throughput can be increased by pairing users enjoying good  channel conditions with users suffering from poor channel conditions.
In \cite{ding2015general}, a transmission framework based on signal alignment was proposed for MIMO NOMA systems.
A suboptimal joint power allocation and precoding design was presented in \cite{minmaxNOMA} for the maximization of the system throughput in multiuser MIMO NOMA single-carrier systems.
Spectral efficiency can be further improved by applying NOMA in multicarrier systems due to the inherent ability of multicarrier systems to exploit multiuser diversity. However, a careful design of power allocation and user scheduling is necessary for multicarrier NOMA (MC-NOMA) systems due to the unavoidable CCI.
In \cite{saito2013non}, the authors demonstrated that MC-NOMA systems  achieve a system throughput gain over conventional multicarrier orthogonal multiple access (MC-OMA) systems for a suboptimal power allocation scheme.
In \cite{OFDM_NOMA_DC}, a suboptimal power allocation algorithm was proposed for the maximization of the weighted system throughput in two-user OFDM based NOMA systems.
The authors of \cite{JointNOMA} proposed a suboptimal joint power and subcarrier allocation algorithm for MC-NOMA systems.
However, since the resource allocation schemes proposed in \cite{saito2013non}\nocite{OFDM_NOMA_DC}--\cite{JointNOMA} are strictly suboptimal, the achievable improvement in spectral efficiency of MC-NOMA systems compared to conventional MC-OMA systems is not clear and the optimal resource allocation design for MC-NOMA systems is still an open problem.

Motivated by the aforementioned observations, we formulate the resource allocation algorithm design for the maximization of the weighted system throughput of MC-NOMA systems as a non-convex optimization problem. The optimal power and subcarrier allocation policy can be obtained by solving the considered problem via a monotonic optimization approach \cite{tuy2000monotonic}\nocite{zhang2013monotonic}--\cite{bjornson2013optimal}. Also, a low-complexity suboptimal algorithm based on successive convex approximation is proposed  and shown to achieve a close-to-optimal system performance.

\vspace*{-1mm}
\section{System Model}
In this section, we present the adopted notation and the considered MC-NOMA system model.
\vspace*{-1mm}
\subsection{Notation}%
We use boldface lower case letters to denote vectors. $\mathbf{a}^T$ denotes the transpose of vector $\mathbf{a}$; $\mathbb{C}$ denotes the set of complex values; $\mathbb{R}$ denotes the set of non-negative real values; $\mathbb{R}^{N\times 1}$ denotes the set of all $N\times 1$ vectors with real entries and $\mathbb{R}^{N\times 1}_{\mathrm{+}}$ denotes the non-negative subset of $\mathbb{R}^{N\times 1}$; $\mathbb{Z}^{N\times 1}$ denotes the set of all $N\times 1$ vectors with integer entries; $\mathbf{a} \le \mathbf{b}$ indicates that $\mathbf{a}$ is component-wise smaller than $\mathbf{b}$; $\abs{\cdot}$ denotes the absolute value of a complex scalar; ${\cal E}\{\cdot\}$ denotes statistical expectation. The circularly symmetric complex Gaussian distribution with mean $w$ and variance $\sigma^2$ is denoted by ${\cal CN}(w,\sigma^2)$; and $\sim$ stands for ``distributed as". $\nabla_{\mathbf{x}} f(\mathbf{x})$ denotes the gradient vector of function $f(\mathbf{x})$ whose components are the partial derivatives of $f(\mathbf{x})$.
\vspace*{-1mm}
\subsection{MC-NOMA System}
\begin{figure}[t]
 \centering
\includegraphics[width=3.5in]{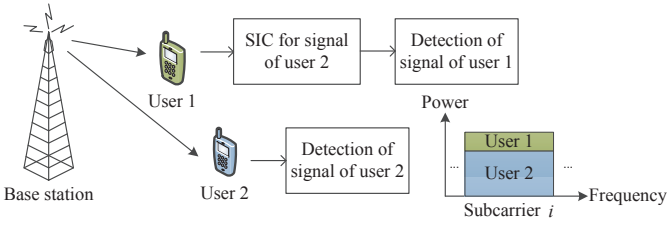} \vspace*{-2mm}
\caption{An MC-NOMA system where two users are multiplexed on subcarrier $i$. The downlink channel quality between user $1$ and the BS is better than that between user $2$ and the BS.
User 1 decodes and removes user $2$'s signal before decoding its own desired signal. The power allocated to user $2$ on subcarrier $i$ is higher than that allocated to user 1.
}
\label{fig:NOMA-model}\vspace*{-3mm}
\end{figure}\vspace*{-0mm}

We consider a downlink MC-NOMA system which consists of a base station (BS) and $K$ downlink users. All transceivers are equipped with a single antenna. The entire frequency band of $W$ Hertz is partitioned into ${N_{\mathrm{F}}}$ orthogonal subcarriers.
In this paper, we assume that each subcarrier is allocated to at most two users to reduce CCI on each subcarrier\footnote{The CCI per subcarrier increases as more users are multiplexed on the same subcarrier which can degrade the system performance.} and to ensure low hardware complexity and low processing delay\footnote{NOMA transmission is enabled by SIC at the receivers. SIC requires demodulation and decoding of the signals intended for other users in addition to the own signal. Thus, hardware complexity and processing delay increase with the number of users multiplexed on the same subcarrier \cite{saito2013non}.}. Each user is equipped with a successive interference canceller, cf. Figure \ref{fig:NOMA-model}.

The received signals at downlink user $m\in \{1,\ldots,K\}$ and downlink user $n\in \{1,\ldots,K\}$ on subcarrier $i \in \{1,\ldots,N_{\mathrm{F}}\}$ are given by \vspace*{-2mm}
\begin{eqnarray}
\hspace*{-5mm}&& y_m^i=\sqrt{p_m^i \rho_m}h_m^i x_m^i + \sqrt{p_n^i \rho_m}h_m^i x_n^i + z_m^i  \,\,\,\, \text{and}\notag \\
\hspace*{-5mm}&& y_n^i=\sqrt{p_n^i \rho_n}h_n^i x_n^i + \sqrt{p_m^i \rho_n}h_n^i x_m^i + z_n^i,
\end{eqnarray}
respectively, where $x_m^i\in\mathbb{C}$ denotes the symbol transmitted from the BS to user $m$ on subcarrier $i$, and we assume ${\cal E}\{\abs{x_m^i}^2\}=1$ without loss of generality.  $p_m^i$ is the transmit power of the signal intended for user $m$ on subcarrier $i$ at the BS. $h_m^i\in\mathbb{C}$ denotes the small scale fading coefficient for the link between the BS and user $m$ on subcarrier $i$. Variable $\rho_m\in\mathbb{R}$ represents the joint effect of path loss and shadowing between the BS and user $m$. $z_m^i\sim{\cal CN}(0,\sigma_{\mathrm{z}_m}^2)$ denotes the complex additive white Gaussian noise (AWGN) on subcarrier $i$ at user $m$. Besides, for the study of optimal resource allocation algorithm design, we assume that the global channel state information (CSI) of all users is perfectly known at the BS.

\vspace*{-1mm}
\section{Problem Formulation}
In this section, we first define the adopted performance measure for the considered MC-NOMA system. Then, we formulate the power and subcarrier allocation problem.
\vspace*{-1mm}
\subsection{Weighted System Throughput}
NOMA systems exploit the power domain for multiple access where different users are served at different power levels.
In particular, for a given subcarrier, a user who enjoys a better downlink channel quality can decode and remove the CCI from a user who has a worse downlink channel quality by employing SIC \cite{ImpactPairNOMA}\nocite{ding2015general,minmaxNOMA,saito2013non,OFDM_NOMA_DC}--\cite{JointNOMA}. Thus, assuming that users $m$ and $n$ are multiplexed on subcarrier $i$ and user $n$ enjoys a better BS-to-user link quality than user $m$ on subcarrier $i$, the instantaneous weighted throughput on subcarrier $i$ is given by \vspace*{-2mm}
\begin{eqnarray} \label{throughput_i}
&&\hspace*{-6mm}U_{m,n}^i(p_{m}^i,p_{n}^i,s_{m,n}^{i})\hspace*{-0.5mm} \notag\\[-1mm]
=&&\hspace*{-6mm}s_{m,n}^{i}\Big[\hspace*{-0.5mm}w_m \hspace*{-0.5mm} \log_2\hspace*{-0.5mm} \Big(1 \hspace*{-1mm}+ \hspace*{-1mm} \frac{H_{m}^i p_{m}^i } {H_{m}^i p_{n}^i \hspace*{-1mm} +\hspace*{-1mm} 1}\Big) \hspace*{-1mm} +\hspace*{-0.5mm} w_n\log_2(1 \hspace*{-1mm}+ \hspace*{-1mm}H_{n}^i p_{n}^i )\Big],
\end{eqnarray}
where $H_{m}^i \hspace*{-0mm} = \hspace*{-0mm} \frac{\rho_m \abs{h_m^i}^2}{\sigma_{\mathrm{z}_m}^2}$, $H_{m}^i \hspace*{-0mm} \le \hspace*{-0mm} H_{n}^i$, and the positive constant $0 \hspace*{-0mm} \le\hspace*{-0mm}  w_m \hspace*{-0mm}\le \hspace*{-0mm}1$ denotes the priority of user $m$ in resource allocation, which is specified in the media access control (MAC) layer to achieve certain fairness objectives.
We note that user $n$ can decode and remove the CCI from user $m$ successfully since $\log_2(1+\frac{H_n^i p_m^i}{H_n^i p_n^i + 1}) \ge \log_2(1+\frac{H_m^i p_m^i}{H_m^i p_n^i + 1})$ when $H_{m}^i \hspace*{-0mm} \le \hspace*{-0mm} H_{n}^i$.
Thus, user $n$'s instantaneous weighted throughput on subcarrier $i$ is $w_n\log_2(1 \hspace*{-1mm}+ \hspace*{-1mm}H_{n}^i p_{n}^i )$. User $m$ cannot perform SIC and regards user $n$'s signal as interference.
Furthermore, $s_{m,n}^{i}$ is the subcarrier allocation indicator which is given by \vspace*{-2mm}
\begin{equation}
s_{m,n}^{i}\hspace*{-0mm}=
\hspace*{-0mm} \begin{cases}
\hspace*{-0mm} 1 & \hspace*{-0mm} \textrm{if user $m$ and user $n$ are multiplexed  }\\
                 & \hspace*{-0mm} \textrm{on subcarrier $i$ with $H_{m}^i \le H_{n}^i$,} \\
\hspace*{-0mm} 0 & \hspace*{-0mm} \textrm{otherwise.}
\end{cases}
\end{equation}

We note that for the case of $m=n$, the instantaneous weighted throughput on subcarrier $i$ in \eqref{throughput_i} becomes \vspace*{-2mm}
\begin{eqnarray} \label{OMA_throughput_i}
\hspace*{-0mm}U_{m,n}^i(p_{m}^i,p_{n}^i,s_{m,n}^{i})\hspace*{-0mm}
=\hspace*{-0mm}s_{m,n}^{i} \hspace*{-0.5mm}w_m \hspace*{-0.5mm} \log_2\hspace*{-0.5mm} \big(1 \hspace*{-1mm}+ \hspace*{-1mm} H_{m}^i (p_{m}^i +p_{n}^i)\ \big).
\end{eqnarray}
In fact, \eqref{OMA_throughput_i} is the instantaneous weighted throughput of subcarrier $i$ for MC-OMA, where $p_{m}^i +p_{n}^i, \forall m=n$, is the transmit power allocated to user $m$ on subcarrier $i$. Therefore, \eqref{throughput_i}  generalizes the instantaneous weighted throughput of conventional MC-OMA systems to MC-NOMA systems.
\vspace*{-0mm}
\subsection{Optimization Problem Formulation}
The system objective is the maximization of the weighted system throughput. The optimal joint power and subcarrier allocation policy is obtained by solving the following optimization problem:\vspace*{-3mm}
\begin{eqnarray} \label{pro}
&&\hspace*{-1mm}\underset{\mathbf{p},\mathbf{s}}{\maxo}\,\, \,\, \notag \sum_{i=1}^{N_{\mathrm{F}}}\sum_{m=1}^{K} \sum_{n=1}^{K} U_{m,n}^i(p_{m}^i,p_{n}^i,s_{m,n}^{i})\\
\notag\mbox{s.t.}
&&\hspace*{-0mm}\mbox{C1: }\overset{N_{\mathrm{F}}}{\underset{i=1}{\sum}} \overset{K}{\underset{m=1}{\sum}} \overset{K}{\underset{n=1}{\sum}} s_{m,n}^i (p_{m}^i+p_{n}^i) \le P_{\mathrm{max}}, \notag\\
&&\hspace*{-0mm}\mbox{C2: } s_{m,n}^i \in \{0,1\},\,\,\ \forall i,m,n, \notag\\
&&\hspace*{-0mm}\mbox{C3: } \overset{K}{\underset{m=1}{\sum}} \overset{K}{\underset{n=1}{\sum}} s_{m,n}^i \le 1,\,\, \forall i,\notag \\
&&\hspace*{-0mm}\mbox{C4: } p_m^i \ge 0, \,\, \forall i,m,
\end{eqnarray}
where $\mathbf{p}\in\mathbb{R}^{{N_{\mathrm{F}}}K \times 1}$ and $\mathbf{s}\in\mathbb{Z}^{{N_{\mathrm{F}}}K^2 \times 1}$ are the collections of optimization variables $p_m^i$ and $s_{m,n}^i$, respectively.
Constraint C1 is a power constraint for the BS with maximum transmit power allowance $P_{\mathrm{max}}$. Constraints C2 and C3 are imposed to guarantee that each subcarrier is allocated to at most two users.
Here, we note that user pairing is performed on each subcarrier.
Constraint C4 is the non-negative transmit power constraint. We note that the joint power and subcarrier allocation for conventional MC-OMA systems is a subcase of our proposed MC-NOMA problem formulation in \eqref{pro}. In fact, for the case of $s_{m,n}^i=1$, $m=n$,  subcarrier $i$ is exclusively allocated to user $m$ and the subcarrier assignment strategy for subcarrier $i$ reduces to the conventional orthogonal assignment. Besides, we note that the condition of $H_{m}^i \hspace*{-0mm} \le \hspace*{-0mm} H_{n}^i$ is implicitly included in the definition of  $U_{m,n}^i(p_{m}^i,p_{n}^i,s_{m,n}^{i})$.

The problem in \eqref{pro} is a mixed combinatorial non-convex problem due to the integer constraint for subcarrier allocation in C2 and the non-convex objective function. In general, there is no systematic approach for solving mixed combinatorial non-convex problems. However, in the next section, we will exploit the monotonicity of the problem in \eqref{pro} to design the optimal resource allocation strategy for the considered system.

\vspace*{-1mm}
\section{Solutions of the Optimization Problem}
In this section, we solve the problem in \eqref{pro} optimally by applying monotonic optimization. Subsequently, a suboptimal scheme is proposed which achieves close-to-optimal performance with a low computational complexity.
\vspace*{-1mm}
\subsection{Monotonic Optimization}
First, we introduce some mathematical preliminaries of monotonic optimization \cite{tuy2000monotonic}\nocite{zhang2013monotonic}--\cite{bjornson2013optimal}.
\begin{Def}[Box]
Given any vector $\mathbf{z}\in\mathbb{R}^{N\times 1}_{\mathrm{+}}$, the hyper rectangle $[\mathbf{0},\mathbf{z}]=\{\mathbf{x}\mid \mathbf{0}\le\mathbf{x}\le\mathbf{z}\}$ is referred to as a box with vertex $\mathbf{z}$.
\end{Def}
\begin{Def}[Normal]
An infinite set $\mathcal{Z} \subset \mathbb{R}^{N\times 1}_{\mathrm{+}}$ is normal if given any element $\mathbf{z} \in \mathcal{Z}$, the box $[\mathbf{0},\mathbf{z}]\subset\mathcal{Z}$.
\end{Def}
\begin{Def}[Polyblock]
Given any finite set $\mathcal{V} \subset \mathbb{R}^{N\times 1}_{\mathrm{+}}$, the union of all boxes $[\mathbf{0},\mathbf{z}]$, $\mathbf{z}\in\mathcal{V}$, is a polyblock with vertex set $\mathcal{V}$.
\end{Def}
\begin{Def}[Projection]
Given any non-empty normal set $\mathcal{Z} \subset \mathbb{R}^{N\times 1}_{\mathrm{+}}$ and any vector $\mathbf{z}\in\mathbb{R}^{N\times 1}_{\mathrm{+}}$, $\Phi (\mathbf{z})$ is the projection of $\mathbf{z}$ onto the boundary of $\mathcal{Z}$, i.e.,  $\Phi\big(\mathbf{z}\big)=\lambda\mathbf{z}$, where $\lambda=\max\{\beta\mid \beta \mathbf{z} \in \mathcal{Z}\}$ and $\beta\in\mathbb{R}_{\mathrm{+}}$.
\end{Def}
\begin{Def} An optimization problem belongs to the class of monotonic optimization problems if it can be represented in the following form: \vspace*{-2mm}
    \begin{eqnarray} \label{MO}
    &&\hspace*{-10mm}\underset{\mathbf{z}}{\maxo} \,\,\,\Psi(\mathbf{z})\notag\\
    &&\hspace*{-10mm}\mbox{s.t.} \hspace*{7mm}\mathbf{z}\in\mathcal{Z},
    \end{eqnarray}
    where $\mathbf{z}$ is the vertex and set $\mathcal{Z}\subset \mathbb{R}^{N\times 1}_{\mathrm{+}}$ is a non-empty normal closed set and function $\Psi(\mathbf{z})$ is an increasing function on $\mathbb{R}^{N\times 1}_{\mathrm{+}}$.
\end{Def}
\vspace*{-1mm}
\subsection{Joint Power and Subcarrier Allocation Algorithm}
To facilitate the presentation of the optimal resource allocation algorithm in the sequel,  we rewrite the weighted throughput of subcarrier $i$ in \eqref{throughput_i} in an equivalent form:  \vspace*{-2mm}
\begin{eqnarray}
&&\hspace*{-6mm}U_{m,n}^i(p_{m}^i,p_{n}^i,s_{m,n}^{i})\hspace*{-0.5mm} \notag\\[-1mm]
=&&\hspace*{-6mm}\hspace*{-0.5mm}w_m \hspace*{-0.5mm} \log_2\hspace*{-0.5mm} \Big(1 \hspace*{-1mm}+ \hspace*{-1mm} \frac{s_{m,n}^{i} H_{m}^i p_{m}^i } {H_{m}^i p_{n}^i \hspace*{-1mm} +\hspace*{-1mm} 1}\Big) \hspace*{-1mm} +\hspace*{-0.5mm} w_n\log_2(1 \hspace*{-1mm}+ \hspace*{-1mm}s_{m,n}^{i}H_{n}^i p_{n}^i ) \notag\\
=&&\hspace*{-6mm}\hspace*{-0.5mm}w_m \hspace*{-0.5mm} \log_2\hspace*{-0.5mm} \Big(1 \hspace*{-1mm}+ \hspace*{-1mm} \frac{H_{m}^i \tilde{p}_{m,n,m}^i } {H_{m}^i \tilde{p}_{m,n,n}^i \hspace*{-1mm} +\hspace*{-1mm} 1}\Big) \hspace*{-1mm} +\hspace*{-0.5mm} w_n\log_2(1 \hspace*{-1mm}+ \hspace*{-1mm}H_{n}^i \tilde{p}_{m,n,n}^i ) \notag \\
=&&\hspace*{-6mm}\log_2(u_{m,n}^i)^{w_m}+ \log_2(v_{m,n}^i)^{w_n},
\end{eqnarray}
where
$\hspace*{-0mm}u_{m,n}^i=\hspace*{-0.5mm}1 +  \frac{H_{m}^i \tilde{p}_{m,n,m}^i } {H_{m}^i \tilde{p}_{m,n,n}^i \hspace*{-0.5mm} +\hspace*{-0.5mm} 1}$, $v_{m,n}^i\hspace*{-0.5mm}=1+ H_{n}^i \tilde{p}_{m,n,n}^i$, and $\tilde{p}_{m,n,m}^i=s_{m,n}^i p_{m}^i$.
 Then, the original problem in \eqref{pro} can be rewritten as \vspace*{-2mm}
\begin{eqnarray} \label{eqv-pro}
\hspace*{-1mm}&&\hspace*{-0mm}\underset{\tilde{\mathbf{p}},\mathbf{s}}{\maxo}\,\, \,\, \notag \sum_{i=1}^{N_{\mathrm{F}}}\sum_{m=1}^{K} \sum_{n=1}^{K} \log_2(u_{m,n}^i)^{w_m}+ \log_2(v_{m,n}^i)^{w_n} \notag \\
\hspace*{-1mm}&&\hspace*{4mm}\mbox{s.t.}  \,\,\mbox{C1: }\overset{N_{\mathrm{F}}}{\underset{i=1}{\sum}} \overset{K}{\underset{m=1}{\sum}} \overset{K}{\underset{n=1}{\sum}} \, \tilde{p}_{m,n,m}^i\hspace*{-0mm}+\hspace*{-0mm}\tilde{p}_{m,n,n}^i \hspace*{-0mm}\le \hspace*{-0mm}P_{\mathrm{max}}, \notag \\
&&\hspace*{9mm}\mbox{C2, C3},  \hspace*{0mm}\mbox{ C4: } \tilde{p}_{m,n,m}^i \ge 0, \,\, \forall m,n,i,
\end{eqnarray}
where
$\tilde{\mathbf{p}}\in\mathbb{R}^{2N_{\mathrm{F}}K^2 \times1}$ is the collection of all $\tilde{p}_{m,n,m}^i$ and $\tilde{p}_{m,n,n}^i$.

Then, we define\vspace*{-2mm}
\begin{equation}
f_{d}(\tilde{\mathbf{p}})\hspace*{-0mm}=
\hspace*{-0mm} \begin{cases}
\hspace*{-0mm} 1 \hspace*{-0mm} + \hspace*{-0mm} H_{m}^i \hspace*{-0mm} (\tilde{p}_{m,n,m}^i \hspace*{-0mm} + \hspace*{-0mm} \tilde{p}_{m,n,n}^i \hspace*{-0mm} ),  &\hspace*{-0mm} d \hspace*{-0mm} = \Delta, \\
\hspace*{-0mm} 1+H_{n}^i \tilde{p}_{m,n,n}^i, & \hspace*{-0mm} d\hspace*{-0mm} = D/2 + \Delta,
\end{cases}
\end{equation}\vspace*{-1mm}
\begin{equation}
\hspace*{-21.5mm}g_{d}(\tilde{\mathbf{p}})\hspace*{-0mm}=
\hspace*{-0mm} \begin{cases}
\hspace*{-0mm} 1+H_{m}^i \tilde{p}_{m,n,n}^i,  & d=\Delta, \\
\hspace*{-0mm} 1, & d=D/2+\Delta,
\end{cases}
\end{equation}
where $\Delta=(i-1)K^2+(m-1)K+n$ and $D=2N_{\mathrm{F}}K^2$.
We further define $\mathbf{z}\hspace*{-1mm}=\hspace*{-1mm}[z_1\hspace*{-0.5mm},\hspace*{-0.5mm}\ldots\hspace*{-0.5mm},\hspace*{-0.5mm}z_D]^T\hspace*{-1.8mm}=\hspace*{-1mm}[u_{1,1}^1\hspace*{-0.5mm}, \hspace*{-0.5mm}\ldots\hspace*{-0.5mm},\hspace*{-0.5mm}u_{K,K}^{N_{\mathrm{F}}}\hspace*{-0.3mm},\hspace*{-0.3mm}v_{1,1}^1\hspace*{-0.5mm},\hspace*{-0.5mm}\ldots\hspace*{-0.5mm},\hspace*{-0.5mm}v_{K,K}^{N_{\mathrm{F}}}]^T$.
Now, the original problem in $\eqref{pro}$ can be written as a monotonic optimization problem as: \vspace*{-4mm}
\begin{eqnarray}\label{MO-pro}
&&\hspace*{-15mm}\underset{\mathbf{z}}{\maxo}\,\, \,\, \notag \sum_{d=1}^{D} \log_2(z_d)^{\mu_d} \\
\mbox{s.t.}
&&\hspace*{-0mm} \mathbf{z}\in\mathcal{Z},
\end{eqnarray}
where $\hspace*{-0mm}\mu_d\hspace*{-0mm}$ is the equivalent user weight for $z_d$, i.e., $\mu_d\hspace*{-1mm}=\hspace*{-1mm}w_m$, $\forall d \in \{1,\ldots,D/2\}$, and $\mu_d=w_n, \forall d \in \{D/2+1,\ldots,D\}$. The feasible set $\mathcal{Z}$ is given by \vspace*{-1mm}
\begin{eqnarray}
\mathcal{Z}\hspace*{-1mm}=\Big\{ \mathbf{z} \mid 1\le z_d \le \frac{f_d(\tilde{\mathbf{p}})}{g_d(\tilde{\mathbf{p}})},  \tilde{\mathbf{p}}\in\mathcal{P}, \mathbf{s}\in\mathcal{S}, \forall d\Big\},
\end{eqnarray}
where $\mathcal{P}$ and $\mathcal{S}$ are the feasible sets spanned by constraints $\mbox{C1, C2, C3}$, and $\mbox{C4}$.

\begin{figure}[t]
 \centering
\includegraphics[width=3in]{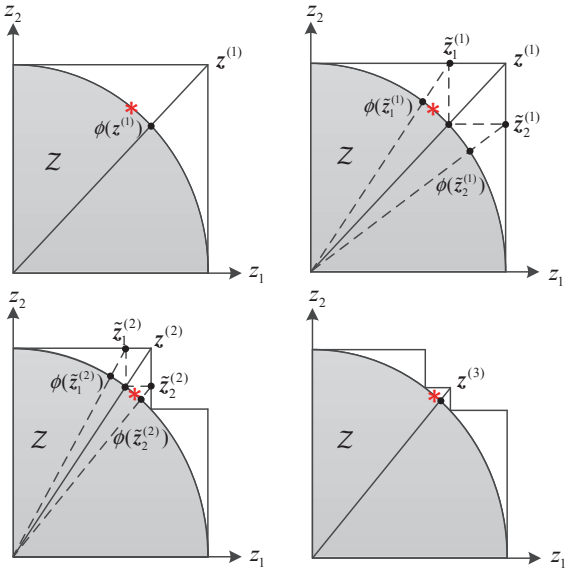} \vspace*{-0mm}
\caption{Illustration of the outer polyblock approximation algorithm for $D=2$. The red star is the optimal point on the boundary of the feasible set $\mathcal{Z}$.} \label{fig:polyblock}\vspace*{-4mm}
\end{figure}

\begin{table}
\begin{algorithm} [H]                    
\caption{Outer Polyblock Approximation Algorithm}          
\label{alg1}                           
\begin{algorithmic} [1]
\small          
\STATE Initialize polyblock $\mathcal{B}^{{(1)}}$ with vertex set $\mathcal{V}^{{(1)}}=\{\mathbf{z}^{{(1)}}\}$ where the elements of $\mathbf{z}^{{(1)}}$ are set as \vspace*{-1mm}
    \begin{eqnarray}
    u_{m,n}^i=1+H_{m}^i P_{\mathrm{max}} \quad\text{and} \quad v_{m,n}^i=1+H_{n}^i P_{\mathrm{max}} \notag
    \end{eqnarray}
\STATE Set error tolerance $\epsilon \ll 1$ and iteration index $k=1$

\REPEAT [Main Loop]
\STATE Construct a smaller polyblock $\mathcal{B}^{(k+1)}$ with vertex set $\mathcal{V}^{{(k+1)}}$ by replacing $\mathbf{z}^{{(k)}}$ with $D$ new vertices $\big\{\tilde{\mathbf{z}}^{{(k)}}_{1}, \ldots,\tilde{\mathbf{z}}^{{(k)}}_{D}\big\}$. The new vertex $\tilde{\mathbf{z}}^{{(k)}}_{d}$, $d\in\{1,\ldots,D\}$, is generated as \vspace*{-1mm}
        \begin{eqnarray}
        \tilde{\mathbf{z}}^{{(k)}}_{d}=\mathbf{z}^{{(k)}}-\Big(z^{{(k)}}_d-\phi_d\big(\mathbf{z}^{{(k)}}\big)\Big)\mathbf{e}_d, \notag
        \end{eqnarray}
        where $\phi_d\big(\mathbf{z}^{{(k)}}\big)$ is the $d$-th element of $\Phi\big(\mathbf{z}^{{(k)}}\big)$ which is obtained by $\textbf{Algorithm 2}$

\STATE  Find $\mathbf{z}^{{(k+1)}}$ as that vertex from $\mathcal{V}^{{(k+1)}}$ whose projection maximizes the objective function of the problem, i.e., \vspace*{-2mm}
        \begin{eqnarray}
        \mathbf{z}^{{(k+1)}}=\underset{\mathbf{z}\in\mathcal{V}^{{(k+1)}}}{\arg \max} \Big\{ \sum_{d=1}^{D} \log_2\big(\phi_d(\mathbf{z})\big)^{\mu_d}\Big\} \notag
        \end{eqnarray}

\STATE  $k=k+1$

\UNTIL $\frac{\norm{\mathbf{z}^{{(k)}}-\Phi(\mathbf{z}^{{(k)}})}}{\norm{\mathbf{z}^{{(k)}}}} \le \epsilon$
\STATE $\mathbf{z}^{*}=\Phi\big(\mathbf{z}^{{(k)}}\big)$ and $\tilde{\mathbf{p}}^*$ is obtained when calculating $\Phi\big(\mathbf{z}^{{(k)}}\big)$
\end{algorithmic}
\end{algorithm}\vspace*{-10mm}
\end{table}
Now, we design a joint power and subcarrier allocation algorithm for solving the monotonic optimization problem in \eqref{MO-pro} based on the outer polyblock approximation approach \cite{tuy2000monotonic}. Since the objective function in \eqref{MO-pro} is a monotonic increasing function, the optimal solution is at the boundary of the feasible set $\mathcal{Z}$ \cite{tuy2000monotonic}\nocite{zhang2013monotonic}--\cite{bjornson2013optimal}. However, the boundary of $\mathcal{Z}$ is unknown. Therefore, we aim to approach the boundary by constructing a sequence of polyblocks. First, we construct a polyblock $\mathcal{B}^{{(1)}}$ that contains the feasible set $\mathcal{Z}$ with vertex set $\mathcal{V}^{{(1)}}$ which includes only one vertex $\mathbf{z}^{{(1)}}$. Then, we construct a smaller polyblock $\mathcal{B}^{{(2)}}$ based on $\mathcal{B}^{{(1)}}$ by replacing $\mathbf{z}^{{(1)}}$ with $D$ new vertices $\tilde{\mathcal{V}}^{{(1)}}=\big\{\tilde{\mathbf{z}}^{{(1)}}_{1}, \ldots,\tilde{\mathbf{z}}^{{(1)}}_{D}\big\}$. The feasible set $\mathcal{Z}$ is still contained in $\mathcal{B}^{{(2)}}$.
The new vertex $\tilde{\mathbf{z}}^{{(1)}}_{d}$ is generated as $\tilde{\mathbf{z}}^{{(1)}}_{d}=\mathbf{z}^{{(1)}}-\Big(z^{{(1)}}_d -\phi_d\big(\mathbf{z}^{{(1)}}\big)\Big)\mathbf{e}_d$, where $\Phi\big(\mathbf{z}^{{(1)}}\big)\in \mathbb{C}^{D \times 1}$ is the projection of $\mathbf{z}^{{(1)}}$ on the feasible set $\mathcal{Z}$, $\phi_d\big(\mathbf{z}^{{(1)}}\big)$ is the $d$-th element of $\Phi\big(\mathbf{z}^{{(1)}}\big)$, and $\mathbf{e}_d$ is a unit vector that has a non-zero element only at index $d$.
Thus, the vertex set $\mathcal{V}^{{(2)}}$ of the newly generated polyblock $\mathcal{B}^{{(2)}}$ is $\mathcal{V}^{{(2)}}=(\mathcal{V}^{{(1)}}-\mathbf{z}^{{(1)}})\cup \tilde{\mathcal{V}}^{{(1)}}$. Then, we choose the optimal vertex from $\mathcal{V}^{{(2)}}$ whose projection maximizes the objective function of the problem in \eqref{MO-pro}, i.e., $\mathbf{z}^{{(2)}}=\underset{\mathbf{z}\in\mathcal{V}^{{(2)}}}{\arg \max} \Big\{ \sum_{d=1}^{D} \log_2\big(\phi_d(\mathbf{z})\big)^{\mu_d}\Big\}$. Similarly, we can repeat the above procedure to construct a smaller polyblock based on $\mathcal{B}^{{(2)}}$ and so on, i.e., $\mathcal{B}^{{(1)}} \supset \mathcal{B}^{{(2)}} \supset \dots \supset \mathcal{Z}$. The algorithm terminates if  $\frac{\norm{\mathbf{z}^{{(k)}}-\Phi(\mathbf{z}^{{(k)}})}}{\norm{\mathbf{z}^{{(k)}}}} \le \epsilon$, where $\epsilon > 0$ is the error tolerance which specifies the accuracy of the approximation. We illustrate the algorithm in Figure \ref{fig:polyblock} for $D=2$. The proposed outer polyblock approximation algorithm is summarized in \textbf{Algorithm 1}. In particular, the vertex $\mathbf{z}^{{(1)}}$ of the initial polyblock $\mathcal{B}^{{(1)}}$ is set by allocating on each subcarrier the maximum transmit power $P_{\mathrm{max}}$ for all users and omitting the CCI. In fact, such intermediate resource allocation policy is infeasible in general. However, the corresponding polyblock contains the feasible set $\mathcal{Z}$ and the algorithm ultimately converges to the optimal point.
\vspace*{-0mm}

The projection of $\mathbf{z}^{{(k)}}$, i.e., $\Phi\big(\mathbf{z}^{{(k)}}\big)=\lambda\mathbf{z}^{{(k)}}$, in \textbf{Algorithm 1}, is obtained by solving \vspace*{-2mm}
\begin{eqnarray} \label{lambda}
\lambda\hspace*{-1mm}&=&\hspace*{-1mm}\max\{\beta\mid \beta \mathbf{z} \in \mathcal{Z}\} \notag \\[-1mm]
\hspace*{-1mm}&=&\hspace*{-1mm} \max\Big\{\beta\mid \beta \le \underset{1\le d \le D}{\min} \,\, \frac{f_d(\tilde{\mathbf{p}})}{z_d^{(k)} g_d(\tilde{\mathbf{p}})}, \tilde{\mathbf{p}}\in\mathcal{P}\Big\}\notag \\[-1mm]
\hspace*{-1mm}&=&\hspace*{-1mm} \underset{\tilde{\mathbf{p}}\in\mathcal{P}}{\max}\underset{1\le d \le D}{\min} \,\, \frac{f_d(\tilde{\mathbf{p}})}{z_d^{(k)} g_d(\tilde{\mathbf{p}})}.
\end{eqnarray}
The problem in \eqref{lambda} is a standard fractional programming problem which can be solved by the Dinkelbach algorithm \cite{JR:fractional} in polynomial time. The algorithm is summarized in \textbf{Algorithm 2}. Specifically, $\tilde{\mathbf{p}}_n^*$ in line 4 is obtained by solving the following convex problem: \vspace*{-2mm}
\begin{eqnarray}\label{max-min-pro}
&&\hspace*{-5mm}\tilde{\mathbf{p}}_n^*=
\underset{\tilde{\mathbf{p}}\in{\mathbf{\mathcal{P}}}}{\arg \max} \,\,\,\, \tau \notag \\[-2mm]
\mbox{s.t.}
&&\hspace*{-5mm} f_d(\tilde{\mathbf{p}})-\lambda_n z_d^{(k)}g_d (\tilde{\mathbf{p}}) \ge \tau, \,\, \forall d \in \{1,\ldots,D\},
\end{eqnarray}
where $\tau$ is an auxiliary variable. Hence, the power allocation policy is obtained when calculating the projection in \textbf{Algorithm 2}.
We note that the convex problem in \eqref{max-min-pro} can be solved by standard convex program solvers such as CVX \cite{website:CVX}.

From the optimal vertex $\mathbf{z}^*$, we can obtain the optimal subcarrier allocation. In particular, we can restore the values of $u_{m,n}^i$ and $v_{m,n}^i$ according to the mapping order of $\mathbf{z}=[z_1,\ldots,z_D]^T=[u_{1,1}^1,\ldots,u_{K,K}^{N_{\mathrm{F}}},v_{1,1}^1,\ldots,v_{K,K}^{N_{\mathrm{F}}}]^T$. Besides, since $u_{m,n}^i$ and $v_{m,n}^i$ are larger than one if users $m$ and $n$ are scheduled on subcarrier $i$, we can obtain the optimal subcarrier allocation policy $\mathbf{s}^*$ as \vspace*{-2mm}
\begin{equation}
s_{m,n}^{i}\hspace*{-0mm}=
\hspace*{-0mm} \begin{cases}
\hspace*{-0mm} 1 & \hspace*{-0mm} u_{m,n}^i > 1, \,\, v_{m,n}^i >1,\\
\hspace*{-0mm} 0 & \hspace*{-0mm} \text{otherwise}.
\end{cases}
\end{equation}

The proposed monotonic optimization based resource allocation algorithm achieves the globally optimal solution. However, its computational complexity grows exponentially with the number of vertices, $D$, used in each iteration. Yet, the performance achieved by the optimal algorithm can serve as a performance upper bound for any suboptimal algorithm. In the following, we propose a suboptimal resource allocation algorithm which has polynomial time computational complexity to strike a balance between complexity and system performance.

\vspace*{-1mm}
\subsection{Suboptimal Solution}
In this section, we propose a low-complexity suboptimal scheme to obtain a local optimal solution for the optimization problem in \eqref{pro}. Since \eqref{eqv-pro} is equivalent to \eqref{pro}, we focus on the solution of the problem in \eqref{eqv-pro}. We note that the product term $\tilde{p}_{m,n,m}^i=s_{m,n}^i p_{m}^i$ in \eqref{eqv-pro} is an obstacle for the design of a computationally efficient resource allocation algorithm. In order to circumvent this difficulty, we adopt the big-M formulation to decompose the product terms \cite{lee2011mixed}. In particular, we impose the following additional constraints: \vspace*{-2mm}
\begin{eqnarray}
&&\hspace*{-4mm}\mbox{C5: } \tilde{p}_{m,n,m}^i \le P_{\mathrm{max}} s_{m,n}^i, \,\, \forall m,n,i, \\
&&\hspace*{-4mm}\mbox{C6: } \tilde{p}_{m,n,m}^i \le p_m^i, \,\, \forall m,n,i,  \\
\hspace*{-4mm}&&\hspace*{-4mm}\mbox{C7: } \tilde{p}_{m,n,m}^i \ge p_m^i \hspace*{-1mm} - \hspace*{-1mm}(1 \hspace*{-1mm} - \hspace*{-0mm}s_{m,n}^i)P_{\mathrm{max}} ,\,\, \forall m,n,i, \,\mbox{and}\\
\hspace*{-4mm}&&\hspace*{-4mm} \mbox{C8: }  \tilde{p}_{m,n,m}^i \ge 0, \,\, \forall m,n,i.
\end{eqnarray}
Besides, the integer constraint C2 in optimization problem \eqref{eqv-pro} is a non-convex constraint. Thus, we rewrite constraint C2 in its equivalent form: \vspace*{-2mm}
\begin{eqnarray}
&&\hspace*{-10mm}\text{C2}\mbox{a: } \overset{{N_{\mathrm{F}}}}{\underset{i=1}{\sum}} \overset{K}{\underset{m=1}{\sum}} \overset{K}{\underset{n=1}{\sum}}s_{m,n}^i -\overset{{N_{\mathrm{F}}}}{\underset{i=1}{\sum}} \overset{K}{\underset{m=1}{\sum}} \overset{K}{\underset{n=1}{\sum}}(s_{m,n}^i)^2 \le 0 \quad \text{and} \\
&&\hspace*{-10mm}\text{C2}\mbox{b: } 0 \le s_{m,n}^i \le 1,\,\, \forall m,n,i.
\end{eqnarray}
\begin{table}[t]
\begin{algorithm} [H]                    
\caption{Projection Algorithm}          
\label{alg1}                           
\begin{algorithmic} [1]
\small          
\STATE Initialize $\lambda_1=0$

\STATE Set error tolerance $\delta \ll 1$ and iteration index $n=1$

\REPEAT
\STATE $\tilde{\mathbf{p}}_n^*=\underset{\tilde{\mathbf{p}}\in{\mathbf{\mathcal{P}}}}{\arg \max} \Big\{\underset{1\le d\le D}{\min}\big\{f_d(\tilde{\mathbf{p}})-\lambda_n z_d^{(k)}g_d (\tilde{\mathbf{p}})\big\}\Big\}$

\STATE $\lambda_{n+1}=\underset{1\le d \le D}{\min}\frac{f_d (\tilde{\mathbf{p}}_n^*)}{z_d^{(k)}g_d (\tilde{\mathbf{p}}_n^*)}$

\STATE $n=n+1$

\UNTIL $\underset{1\le d\le D}{\min}\big\{f_d(\tilde{\mathbf{p}}_{n-1}^*)-\lambda_{n} z_d^{(k)}g_d (\tilde{\mathbf{p}}_{n-1}^*)\big\} \le \delta$
\STATE The projection is $\Phi\big(\mathbf{z}^{{(k)}}\big)=\lambda_{n} \mathbf{z}^{{(k)}}$ and $\tilde{\mathbf{p}}_{n-1}^*$ is the corresponding resource allocation policy.

\end{algorithmic}
\end{algorithm}\vspace*{-10mm}
\end{table}
Now, optimization variables $s_{m,n}^i$ are continuous values between zero and one. However, constraint $\mbox{C2a}$ is the difference of two convex functions which is known as a reverse convex function \cite{che2014Joint}\nocite{ng2015power}--\cite{dinh2010local}. In order to handle constraint $\mbox{C2a}$, we reformulate the problem in \eqref{eqv-pro} as \vspace*{-2mm}
\begin{eqnarray} \label{penalty-pro}
\hspace*{-5mm}&&\hspace*{-5mm}\underset{\tilde{\mathbf{p}},\mathbf{s}}{\mino}\,\, \,\, \notag \sum_{i=1}^{{N_{\mathrm{F}}}}\sum_{m=1}^{K} \sum_{n=1}^{K} -\log_2(u_{m,n}^i)^{w_m} -\log_2(v_{m,n}^i)^{w_n} \notag \\
&&\hspace*{11mm}+ \eta\Big(\overset{{N_{\mathrm{F}}}}{\underset{i=1}{\sum}} \overset{K}{\underset{m=1}{\sum}} \overset{K}{\underset{n=1}{\sum}}s_{m,n}^i -\overset{{N_{\mathrm{F}}}}{\underset{i=1}{\sum}} \overset{K}{\underset{m=1}{\sum}} \overset{K}{\underset{n=1}{\sum}}(s_{m,n}^i)^2\Big) \notag \\
\hspace*{-1mm}&&\hspace*{4mm}\mbox{s.t.}  \hspace*{7mm}\,\,\mbox{C1}, \text{C2}\mbox{b}, \mbox{C3--C8},
\end{eqnarray}
where $\eta \gg 1$ is a large constant which acts as a penalty factor to penalize the objective function for any $s_{m,n}^i$ that is not equal to $0$ or $1$. It can be shown that \eqref{penalty-pro} and \eqref{eqv-pro} are equivalent for $\eta \gg 1$ \cite{che2014Joint,ng2015power}. The resulting optimization problem in \eqref{penalty-pro} is still non-convex because of the objective function. To facilitate the presentation, we rewrite the problem in \eqref{penalty-pro} as \vspace*{-2mm}
\begin{eqnarray}\label{dc-penalty-pro}
\hspace*{-1mm}&&\hspace*{-0mm}\underset{\tilde{\mathbf{p}},\mathbf{s}}{\mino}\,\, \,\, F(\tilde{\mathbf{p}})-G(\tilde{\mathbf{p}})+\eta(H(\mathbf{s})-M(\mathbf{s})) \notag \\
\hspace*{-1mm}&&\hspace*{4mm}\mbox{s.t.}  \hspace*{7mm}\,\,\mbox{C1}, \text{C2}\mbox{b}, \mbox{C3--C8},
\end{eqnarray}
where \vspace*{-2mm}
\begin{eqnarray}
\hspace*{-7mm}F(\tilde{\mathbf{p}})\hspace*{-3mm}&=&\hspace*{-3mm}\sum_{i=1}^{{N_{\mathrm{F}}}}\sum_{m=1}^{K} \sum_{n=1}^{K} -w_m \hspace*{-0.5mm} \log_2\hspace*{-0.5mm} (1 \hspace*{-1mm}+ \hspace*{-1mm} H_{m}^i (\tilde{p}_{m,n,m}^i + \tilde{p}_{m,n,n}^i))  \hspace*{-1mm} \notag \\
\hspace*{-5mm}&-&\hspace*{-3mm} w_n\log_2(1 \hspace*{-1mm}+ \hspace*{-1mm}H_{n}^i \tilde{p}_{m,n,n}^i ),
\end{eqnarray}\vspace*{-6mm}
\begin{eqnarray}
\hspace*{-20mm}G(\tilde{\mathbf{p}})\hspace*{-3mm}&=&\hspace*{-3mm}\sum_{i=1}^{{N_{\mathrm{F}}}}\sum_{m=1}^{K} \sum_{n=1}^{K}-w_m \hspace*{-0.5mm} \log_2\hspace*{0mm} (1+ H_{m}^i \tilde{p}_{m,n,n}^i ),
\end{eqnarray}\vspace*{-4mm}
\begin{eqnarray}
\hspace*{-6mm}H(\mathbf{s}) \hspace*{-3mm}&=&\hspace*{-3mm} \overset{{N_{\mathrm{F}}}}{\underset{i=1}{\sum}} \overset{K}{\underset{m=1}{\sum}} \overset{K}{\underset{n=1}{\sum}}s_{m,n}^i, \text{and} \,\,\, M(\mathbf{s}) \hspace*{-1mm}=\hspace*{-1mm}\overset{{N_{\mathrm{F}}}}{\underset{i=1}{\sum}} \overset{K}{\underset{m=1}{\sum}} \overset{K}{\underset{n=1}{\sum}}(s_{m,n}^i)^2.
\end{eqnarray}
We note that $F(\tilde{\mathbf{p}})$, $G(\tilde{\mathbf{p}})$, $H(\mathbf{s})$, and $M(\mathbf{s})$ are convex functions and the problem in \eqref{dc-penalty-pro} belongs to the class of difference of convex (d.c.) function programming. As a result, we can apply successive convex approximation \cite{dinh2010local} to obtain a local optimal solution of \eqref{dc-penalty-pro}.
Since $G(\tilde{\mathbf{p}})$ and $M(\mathbf{s})$ are differentiable convex functions, for any feasible point $\tilde{\mathbf{p}}^{(k)}$ and $\mathbf{s}^{(k)}$, we have the following inequalities \vspace*{-2mm}
\begin{eqnarray}\label{ineq1}
G(\tilde{\mathbf{p}}) &\ge& G(\tilde{\mathbf{p}}^{(k)}) +\nabla_{\tilde{\mathbf{p}}} G(\tilde{\mathbf{p}}^{(k)})^T(\tilde{\mathbf{p}}-\tilde{\mathbf{p}}^{(k)}) \,\,\, \text{and}\\
\label{ineq2}M(\mathbf{s}) &\ge& M(\mathbf{s}^{(k)}) +\nabla_{\mathbf{s}} M(\mathbf{s}^{(k)})^T(\mathbf{s}-\mathbf{s}^{(k)}),
\end{eqnarray}
where the right hand sides of \eqref{ineq1} and \eqref{ineq2} are affine functions and represent the global underestimation of $G(\tilde{\mathbf{p}})$ and $M(\mathbf{s})$, respectively.
\begin{table}
\begin{algorithm} [H]                    
\caption{Successive Convex Approximation}          
\label{alg1}                           
\begin{algorithmic} [1]
\small          
\STATE Initialize the maximum number of iterations $I_{\mathrm{max}}$, penalty factor $\eta \gg 1$, iteration index $k=1$, and initial point $\tilde{\mathbf{p}}^{(1)}$ and $\mathbf{s}^{(1)}$

\REPEAT
\STATE Solve \eqref{dc} for a given $\tilde{\mathbf{p}}^{(k)}$ and $\mathbf{s}^{(k)}$ and store the intermediate resource allocation policy $\{\tilde{\mathbf{p}}, \mathbf{s}\}$

\STATE Set $k=k+1$ and $\tilde{\mathbf{p}}^{(k)}=\tilde{\mathbf{p}}$ and $\mathbf{s}^{(k)}=\mathbf{s}$

\UNTIL convergence or $k=I_{\mathrm{max}}$

\STATE $\tilde{\mathbf{p}}^{*}=\tilde{\mathbf{p}}^{(k)}$ and $\mathbf{s}^{*}=\mathbf{s}^{(k)}$

\end{algorithmic}
\end{algorithm}\vspace*{-10mm}
\end{table}
Therefore, for any given $\tilde{\mathbf{p}}^{(k)}$ and $\mathbf{s}^{(k)}$, we can obtain an upper bound for \eqref{dc-penalty-pro} by solving the following convex optimization problem: \vspace*{-2mm}
\begin{eqnarray}\label{dc}
\hspace*{-1mm}&&\hspace*{-0mm}\underset{\tilde{\mathbf{p}},\mathbf{s}}{\mino}\,\, \,\, F(\tilde{\mathbf{p}})-G(\tilde{\mathbf{p}}^{(k)}) -\nabla_{\tilde{\mathbf{p}}} G(\tilde{\mathbf{p}}^{(k)})^T(\tilde{\mathbf{p}}-\tilde{\mathbf{p}}^{(k)}) \notag \\
\hspace*{-1mm}&&\hspace*{14mm}+\eta\big(H(\mathbf{s})-M(\mathbf{s}^{(k)}) -\nabla_{\mathbf{s}} M(\mathbf{s}^{(k)})^T(\mathbf{s}-\mathbf{s}^{(k)})\big) \notag \\
\hspace*{-1mm}&&\hspace*{4mm}\mbox{s.t.}  \hspace*{7mm}\,\,\mbox{C1}, \text{C2}\mbox{b}, \mbox{C3--C8},
\end{eqnarray}
where \vspace*{-3mm}
\begin{eqnarray}
\hspace*{-3.5mm}\nabla_{{\mathbf{s}}} M( \hspace*{-0.5mm} \mathbf{s}^{(k)} \hspace*{-0.5mm} )^T \hspace*{-0.5mm} (\mathbf{s}\hspace*{-0.5mm} - \hspace*{-0.5mm}\mathbf{s}^{(k)}\hspace*{-0.5mm} )\hspace*{-3mm}&=&\hspace*{-3mm}\overset{{N_{\mathrm{F}}}}{\underset{i=1}{\sum}} \overset{K}{\underset{m=1}{\sum}} \overset{K}{\underset{n=1}{\sum}}2s_{m,n}^{i (k)}(s_{m,n}^{i}-s_{m,n}^{i (k)}) \,\,\, \text{and} \notag \\
\nabla_{\tilde{\mathbf{p}}}  G( \hspace*{-0.5mm} \tilde{\mathbf{p}}^{(k)} \hspace*{-0.5mm} )^T \hspace*{-0.5mm} (\tilde{\mathbf{p}} \hspace*{-0.5mm} - \hspace*{-0.5mm}\tilde{\mathbf{p}}^{(k)}\hspace*{-0.5mm} )\hspace*{-3mm}&=&\hspace*{-3mm}\overset{{N_{\mathrm{F}}}}{\underset{i=1}{\sum}} \hspace*{-0.5mm}\overset{K}{\underset{m=1}{\sum}}\overset{K}{\underset{n=1}{\sum}} \hspace*{-0.5mm} - \hspace*{-0.5mm} \frac{w_m \hspace*{-0.5mm} H_m^i \hspace*{-0.5mm} (\tilde{p}_{m,n,n}^{i} \hspace*{-1mm} - \hspace*{-0.5mm} \tilde{p}_{m,n,n}^{i(k)})}{(1 \hspace*{-0.5mm} + \hspace*{-0.5mm} H_m^i\tilde{p}_{m,n,n}^{i(k)})\ln2}. \notag
\end{eqnarray}
Then, we employ an iterative algorithm to tighten the obtained upper bound as summarized in \textbf{Algorithm 3}.
The convex problem in \eqref{dc} can be solved efficiently by standard convex program solvers such as CVX \cite{website:CVX}.
By solving the convex upper bound problem in \eqref{dc}, the proposed iterative scheme generates a sequence of feasible solutions $\tilde{\mathbf{p}}^{(k+1)}$ and $\mathbf{s}^{(k+1)}$ successively. The proposed suboptimal iterative algorithm converges to a local optimal solution of \eqref{dc} with polynomial time computational complexity \cite{dinh2010local}.

\vspace*{-1mm}
\section{Simulation Results}
In this section, we investigate the performance of the proposed resource allocation scheme through simulations.
A single cell with two ring-shaped boundary regions is considered. The outer boundary and the inner boundary have radii of $30$ meters and $600$ meters, respectively. The $K$ downlink users are randomly and uniformly distributed between the inner and the outer boundary.
The BS is located at the center of the cell. The number of subcarriers is set to ${N_{\mathrm{F}}}=64$ with a carrier center frequency of $2.5$ GHz and a system bandwidth of $W=5$ MHz. Hence, each subcarrier has a bandwidth of $78$ kHz.
The maximum total transmit power of the BS is $P_{\mathrm{max}}$. The noise power at user $m$ is $\sigma_{\mathrm{z}_m}^2=-128 \text{ dBm}$ on each subcarrier. For the weight of the users, we choose the normalized distance between the users and the BS, i.e., $w_m=\frac{l_m}{\underset{i}{\max}\{l_i\}}$, where $l_m$ is the distance from user $m$ to the BS\footnote{The weights are chosen to ensure resource allocation fairness, especially for the cell edge users which suffer from poor channel conditions. Other fairness strategies can be applied according to the preferences of the system operator, of course.}. The penalty term $\eta$ for the proposed suboptimal algorithm is set to $10 \log_2(1+\frac{P_{\mathrm{max}}}{\sigma_{\mathrm{z}_m}^2})$. The 3GPP path loss model is used with path loss exponent $3.6$ \cite{3GPPchannel}. The small-scale fading of the channel between the BS and the users is modeled as independent and identically distributed Rayleigh fading. The results shown in the following sections were averaged over different realizations of both path loss and multipath fading.

\vspace*{-1mm}
\subsection{Average System Throughput vs. Maximum Transmit Power}
\begin{figure}[t]
 \centering\vspace*{-5mm}
\includegraphics[width=3.4in]{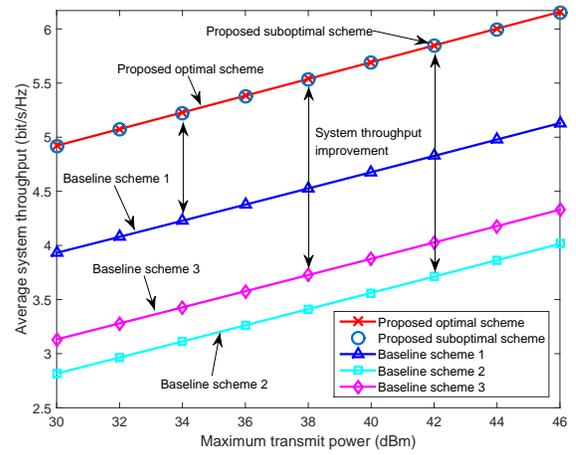} \vspace*{-3mm}
\caption{Average system throughput (bits/s/Hz) vs. the maximum transmit power at the BS (dBm), $P_{\mathrm{max}}$, for different resource allocation schemes and $K=6$ users. The double-sided arrows indicate the performance gains of the proposed optimal scheme compared to the baseline schemes.} \label{fig:wsr_vs_power}\vspace*{-5mm}
\end{figure}

In Figure \ref{fig:wsr_vs_power}, we investigate the average system throughput versus (vs.) the maximum transmit power at the BS, $P_{\mathrm{max}}$, for $K=6$ users. As can be observed, the average system throughput increases monotonically with the maximum transmit power $P_{\mathrm{max}}$ since the received signal-to-interference-plus-noise ratio (SINR) at the users can always be improved by allocating additional available transmit power optimally by solving the problem in \eqref{pro}.
Besides, it can be observed from Figure \ref{fig:wsr_vs_power} that the proposed suboptimal scheme closely approaches the performance of the proposed optimal power and subcarrier allocation scheme.
For comparison, Figure \ref{fig:wsr_vs_power} also shows the average system throughput of three baseline schemes.
For baseline scheme $1$, we adopt the suboptimal joint power and subcarrier allocation for MC-NOMA which was proposed in \cite{JointNOMA}.
For baseline scheme $2$, the user pair on each subcarrier is randomly selected and we optimize the transmit power $p_m^i$ subject to constraints C1-C4 as in \eqref{pro}.
For baseline scheme  $3$, we consider the conventional MC-OMA scheme where each subcarrier can only be allocated to at most one user. Then, we optimize the transmit power of the users and the subcarrier allocation policy to maximize the system throughput given the maximum transmit power allowance $P_{\mathrm{max}}$ at the BS.
The average system throughputs of all baseline schemes are substantially lower than those of the proposed optimal and suboptimal schemes.
In particular, baseline schemes $1$ and $2$ achieve a lower average system throughput compared to the proposed optimal scheme due to their non-optimality in power and subcarrier allocation. For the case of $P_{\mathrm{max}}=46$ dBm, the proposed optimal scheme achieves roughly a $20\%$ and $56\%$ higher average system throughput than baseline schemes $1$ and $2$, respectively. The proposed optimal and suboptimal schemes utilize the available transmit power efficiently. In particular, it can be observed from Figure \ref{fig:wsr_vs_power} that for a given target system throughput, the proposed schemes enable power reductions of more than $10$ dB compared to the baseline schemes. Also, baseline scheme $3$ achieves a lower average system throughput compared to the proposed schemes and baseline scheme 1 since for MC-NOMA the spectrum resource is underutilized due to the orthogonal subcarrier assignment.

\vspace*{-1.4mm}
\subsection{Average System Throughput vs. Number of Users}
\begin{figure}[t]
 \centering\vspace*{-5mm}
\includegraphics[width=3.4in]{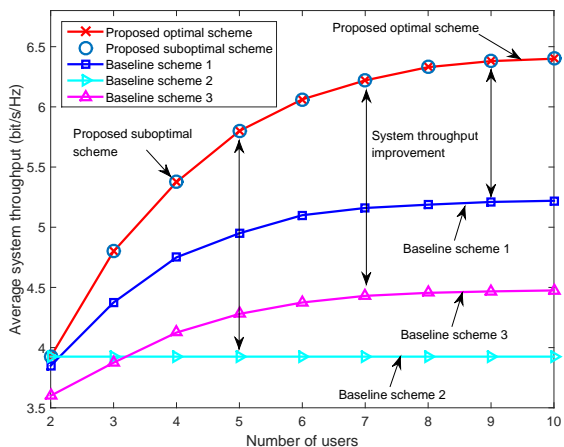} \vspace*{-3mm}
\caption{Average system throughput (bits/s/Hz) vs. the number of users for different resource allocation schemes with $P_{\mathrm{max}}=45$ dBm. The double-sided arrows indicate the performance gains of the proposed optimal scheme compared to the baseline schemes.}\label{fig:wsr_vs_usernum}\vspace*{-5mm}
\end{figure}
In Figure \ref{fig:wsr_vs_usernum}, we investigate the average system throughput vs. the number of users for a maximum transmit power of $P_{\mathrm{max}}=45$ dBm. As can be observed, the average system throughput for the proposed optimal/suboptimal schemes and baseline schemes  1 and 3 increase with the number of users since these schemes are able to exploit multiuser diversity. On the other hand, baseline scheme  $2$ is insensitive to the number of users due to its random scheduling policy. Besides, it can be observed from Figure \ref{fig:wsr_vs_usernum} that the average system throughput of the proposed optimal and suboptimal schemes grows faster with an increasing number of users than that of baseline schemes  $1$ and $3$. In fact, since the proposed MC-NOMA scheme exploits not only the frequency domain but also the power domain for multiple access, more degrees of freedom are available in MC-NOMA systems for user selection and power allocation. Thus, both the proposed optimal scheme and baseline scheme  $1$ achieve a higher system throughput than the MC-OMA system in baseline scheme $3$.
On the other hand, the proposed optimal scheme always achieves a higher system throughput than baseline scheme $1$ due to its optimal power and subcarrier allocation.
We note that the proposed suboptimal scheme achieves a similar performance as the proposed optimal scheme, even for relatively large numbers of users.

\vspace*{-1mm}
\section{Conclusion}
In this paper, we studied the optimal joint power and subcarrier allocation policy for MC-NOMA systems. The resource allocation algorithm design was first formulated as a non-convex optimization problem with the objective to maximize the weighted system throughput. The proposed resource allocation problem was then solved optimally by using monotonic optimization. Besides, a low-complexity suboptimal scheme was also proposed and shown to achieve a close-to-optimal performance. Simulation results unveiled that the proposed MC-NOMA achieves a significant improvement in system performance compared to conventional MC-OMA. Furthermore, our results also showed the importance of efficient resource allocation optimization in NOMA systems.

\vspace*{-1mm}


\end{document}